# Interfacial spintronic THz emission

*Piyush Agarwal, Rohit Medwal, Keynesh Dongol, John Rex Mohan, Yingshu Yang, Hironori Asada, Yasuhiro Fukuma, Ranjan Singh\**

P. Agarwal, K. Dongol, Y. Yang, Prof. R. Singh
Division of Physics and Applied Physics, School of Physical and Mathematical Sciences, Nanyang Technological University, 21 Nanyang Link, Singapore 637371, Singapore
E-mail: ranjans@ntu.edu.sg

P. Agarwal, K. Dongol, Prof. R. Singh
Center for Disruptive Photonic Technologies, The Photonics Institute, Nanyang Technological University, Singapore 639798, Singapore
E-mail: ranjans@ntu.edu.sg

Dr. R. Medwal
Department of Physics, Indian Institute of Technology Kanpur, Uttar Pradesh 208016, India

Prof. H. Asada
Department of Electronic Devices and Engineering, Graduate School of Science and Engineering, Yamaguchi University, Ube 755-8611, Japan

J. R. Mohan, Prof. Y. Fukuma
Department of Physics and Information Technology, Faculty of Computer Science and System Engineering, Kyushu Institute of Technology, Iizuka 820-8502, Japan

**Abstract**

The broken inversion symmetry at the ferromagnet (FM)/heavy-metal (HM) interface leads to spin-dependent degeneracy of the energy band, forming spin-polarized surface states. As a result, the interface serves as an effective medium for converting spin accumulation into two-dimensional charge current through the inverse Rashba-Edelstein effect. Exploring and assessing this spin-to-charge conversion (SCC) phenomenon at the FM/HM interface could offer a promising avenue to surpass the presumed limits of SCC in bulk HM layers. We utilize spintronic heterostructures as a platform to measure the spin-to-charge conversion (SCC) experienced by photoexcited spin currents. These heterostructures emit terahertz electric field when illuminated by femtosecond laser pulses, enabling us to quantitatively assess the ultrafast SCC process. Our results demonstrate a robust interfacial spin-to-charge conversion ($i$SCC) within a synthetic antiferromagnetic heterostructure, specifically for the NiFe/Ru/NiFe configuration, by isolating the SCC contribution originating from the interface itself, separate from the bulk heavy-metal (HM) region. Moreover, the $i$SCC at the NiFe/Ru interface is discovered to be approximately 27% of the strength observed in the highest spin-Hall conducting heavy-metal, Pt. Our results thus highlight the significance of interfacial engineering as a promising pathway for achieving efficient ultrafast spintronic devices.

**Introduction**

Driving electronic spin[1] with femtosecond laser excitation[2,3] has opened a route for generating spin current in FM/HM heterostructure at sub-picosecond timescales[2,4]. The spin current experience an inverse spin-Hall effect(ISHE)[5,6] within the device to produce detectable electrical pulses $(j_c)$[7–10], suggesting the underlying spin relaxation[11,12] to play a pivotal role in the ultrafast spintronics applications[13,14]. Giant enhancement of ISHE in spintronic devices has thus become significant where the spin-to-charge conversion(SCC) has approached saturation due to the limited library of HM, emphasizing only a few materials like Platinum and Tungsten[6]. However, besides the SCC in the bulk HM layer[15], the generated spin current also undergoes SCC at the FM/HM interface[16–18] and co-exists simultaneously. As a result, the interactions between the spins and the interfacial surface states[19–22] have recently gained considerable interest and emerged as a strong candidate for efficient SCC[23–25]. Several supporting studies performed under equilibrium or quasi-equilibrium conditions have exhibited an effective interfacial SCC[21,26], but their influence during non-equilibrium excitation remains underexplored.

The broken inversion symmetry across the interface provides a platform for the SCC through the inverse Rashba-Edelstein effect (IREE)[21,26–31]. The effect further enhances upon utilizing the surface states of heavy metals[8,32]. Attempts have been made to estimate the ultrafast SCC[33,34] using THz emission from heterostructures[35,36], yet an explicit deconvolution of the SCC between the FM/HM interface and bulk HM could not be determined. Here, we quantify the inherent interfacial effects through terahertz emission. In order to highlight the interfacial effect in spin-to-charge conversion (SCC), a trilayer configuration of the heterostructure with a lower spin-Hall conductive heavy metal, specifically NiFe/Ru/NiFe, is employed. Within this synthetic antiferromagnetic heterostructure, the Ruderman–Kittel–Kasuya–Yosida (RKKY) interaction facilitates coupling of the localized *d*-orbital spins between the two ferromagnetic (FM) layers across the interlayer. This allows for independent magnetization control of each ferromagnet within the system. By applying symmetry arguments, we can establish a correlation between the SCC observed in the NiFe/Ru/NiFe and NiFe/Ru configurations. This enables us to discern and differentiate the quantitative contributions that stem from the respective interfaces. The SCC originating from the bulk ruthenium is determined to be negligible in comparison to the $i$SCC at the NiFe/Ru(1nm) interface. Moreover, in terms of absolute magnitude, the $i$SCC at the NiFe/Ru interface is found to reach approximately 27% when compared to the combined SCC from the NiFe/Pt interface and bulk Pt in NiFe/Pt(2nm). The findings provide strong evidence of a substantial SCC at the NiFe/Ru interface, underscoring the robust contribution of $i$SCC in spintronic heterostructures. As such, the significance of $i$SCC has been previously underestimated, emphasizing its potential impact on future spintronic device design and functionality. Thus, by uncovering the significant potential of $i$SCC, we propose interfacial engineering as a crucial pathway for advancing the quest to achieve the highest SCC and enhance emerging ultrafast spintronic devices. Furthermore, manipulating and optimizing the interfaces unlock the full potential of SCC, which is essential for the development of future spintronic technologies.

**Results and Discussion**

The presence of a spin-polarized surface resulting from the broken inversion symmetry within the FM/HM heterostructure is identified as the underlying cause of the inverse Rashba-Edelstein effect (IREE). Simultaneously, the bulk HM layer and bulk FM layer contribute to the inverse spin-Hall effect (ISHE) and anomalous Hall effect (AHE), respectively [37,38]. Together, these phenomena provide simultaneous and distinct sources of SCC in the FM/HM bilayer[16]; however, the bulk HM has restricted an explicit quantification of the contribution from the FM/HM interface[39]. Due to that, although IREE has often been explored from the perspective of equilibrium excitation, a clear and inherent contribution during non-equilibrium photoexcitation of the heterostructure has remained elusive.

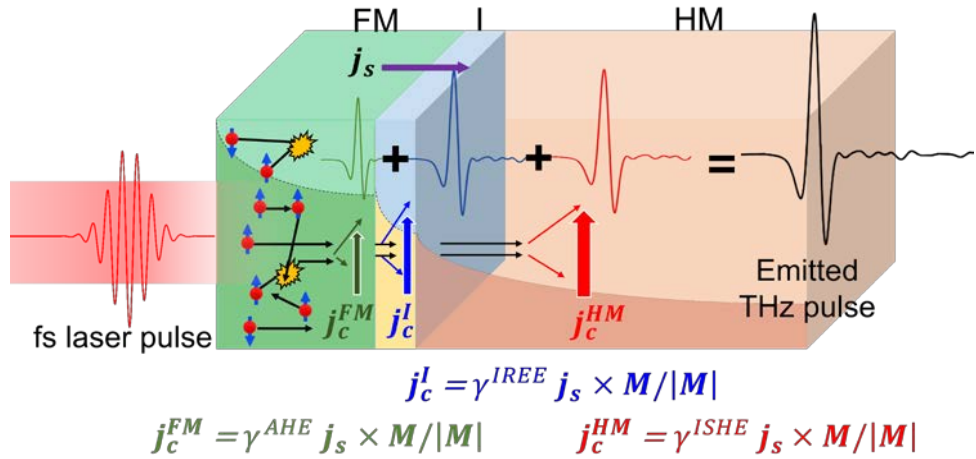

**Figure 1: Role of interface and bulk in spintronic terahertz emission from FM/HM heterostructure -** Terahertz emission from a prototypical spintronic heterostructure, FM/HM, upon femtosecond laser photoexcitation. The schematic depicts the spin-to-charge conversion at the FM/HM interface (shown in blue color), bulk HM (shown in red color), and bulk FM (shown in dark green color) while $\gamma^{IREE}, \gamma^{ISHE}, \gamma^{AHE}$ account for the interfacial and bulk SCC in HM and FM, respectively. The superdiffusive spin current, $\boldsymbol{j_s}$ yields charge current $\boldsymbol{j_c} = \boldsymbol{j_c^I} + \boldsymbol{j_c^{HM}} + \boldsymbol{j_c^{FM}}$.

As shown in **Figure 1**, we investigate the SCC in spintronic heterostructures, which upon femtosecond laser illumination, yield terahertz electric fields given by $\boldsymbol{E}(t) = \partial \boldsymbol{j_c}(t)/\partial t$ [7,8] and $j_c$ denotes the ultrafast charge current[7–9,40]. Owing to the cumulative contributions from the IREE, ISHE, and AHE, $j_c = \gamma^{total} j_s \times M/|M|$[7,16], where $\gamma^{total} = \gamma^{IREE} + \gamma^{ISHE} + \gamma^{AHE}$ and $\gamma^{IREE}$ ($\gamma^{ISHE}$) accounts for the SCC in the FM/HM interface and bulk HM. $j_s$ denotes the superdiffusive spin transport originating from the FM upon photoexcitation[41–43], and $M/|M|$ is the magnetization unit vector of FM. $\gamma^{AHE}$ accounts for the SCC caused by the anomalous Hall effect within the FM which is responsible for the THz emission from the bare ferromagnet[37,38].

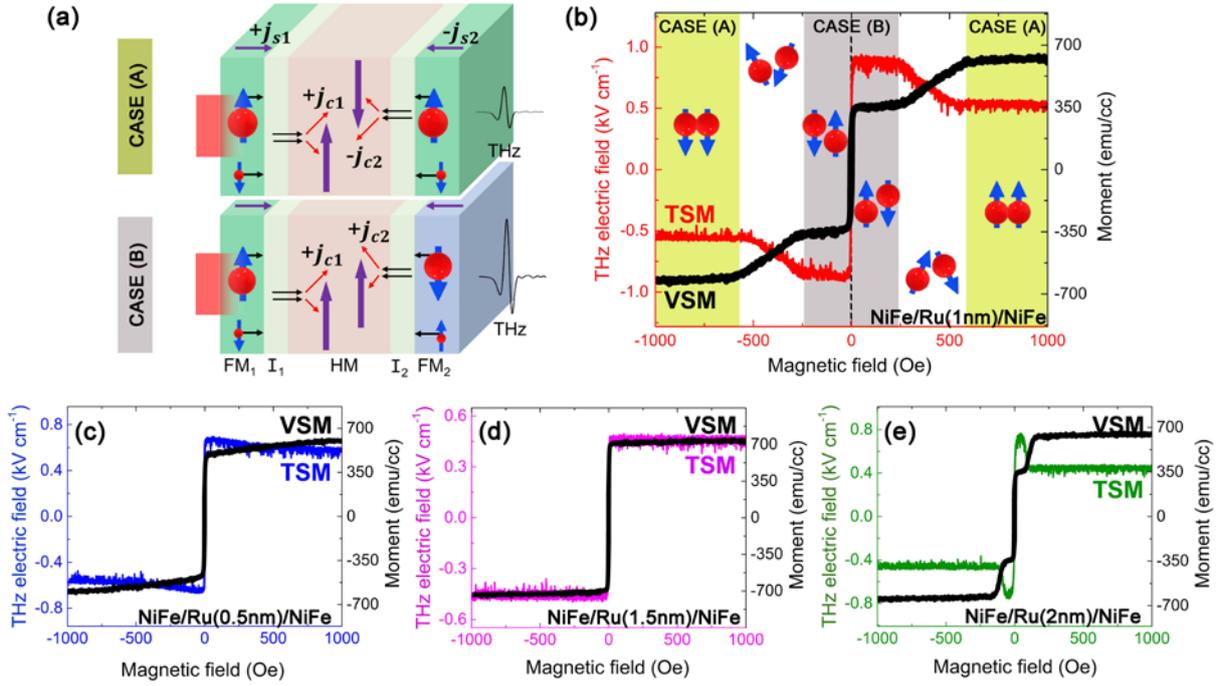

**Figure 2: Terahertz spintronic magnetometer (THz-H hysteresis) based measurements of synthetic antiferromagnetic (SAF) heterostructure -** (a) Schematic diagram depicts spin transport and SCC in exchange-coupled ferromagnets. Attenuated (Enhanced) THz emission is observed when the magnetization in ferromagnets is aligned parallel (anti-parallel) to each other, as shown in Case A (Case B). (b-e) Comparison of the THz-H hysteresis (taken from THz spintronic magnetometer, TSM) and M-H hysteresis (taken from Vibration spintronic magnetometer, VSM) in Qz/NiFe(3nm)/Ru(*x*)/NiFe(3 nm), for (b) *x*=1nm, (c) *x*=0.5nm, (d) *x*=1.5nm, and (e) *x*=2nm. (b) A pair of spin is used to indicate the relative magnetization of the two ferromagnets.

For highlighting the potential of interfacial contribution, Ruthenium(Ru) is chosen as an HM due to its low spin-Hall conductivity, such that the contribution from the bulk HM does not conceal the interfacial effect. In addition, Ru also supports the formation of synthetic antiferromagnets (SAF) states in $FM_1/Ru/FM_2$ heterostructure and allows for active control of the relative magnetization in the two ferromagnets[44]. Due to the tailored magnetization of FM, the transport up-spin and down-spin experience magnetic-field controlled mobility across the FM/HM interface[45]. As a result, the SCC at the interface can be modulated, thus making the SAF heterostructure a preferred system to investigate the interfacial phenomenon.

**Figure 2(a)** illustrates two distinct states of synthetic antiferromagnetic configuration at different magnetic fields. In Case A, the magnetizations of $FM_1$ and $FM_2$ are aligned parallel to each other, producing charge current, $\boldsymbol{j_{c1}}$ from $FM_1 = \gamma_1^{total}\boldsymbol{j_{s1}} \times \boldsymbol{M}/|\boldsymbol{M}|$ and $\boldsymbol{j_{c2}}$ from $FM_2 = \gamma_2^{total}(-\boldsymbol{j_{s2}}) \times \boldsymbol{M}/|\boldsymbol{M}|$. The difference in spin current direction leads to THz emission with the opposite phase, which destructively interferes to emit attenuated radiation. In addition, Case B depicts the configuration when the magnetization of $FM_1$ and $FM_2$ aligns in an anti-parallel direction and produces $\boldsymbol{j_{c1}}$ from $FM_1 = \gamma_1^{total}\boldsymbol{j_{s1}} \times \boldsymbol{M}/|\boldsymbol{M}|$ and $\boldsymbol{j_{c2}}$ from $FM_2 = \gamma_2^{total}(-\boldsymbol{j_{s2}}) \times (-\boldsymbol{M})/|\boldsymbol{M}|$. Due to the anti-parallel spins, Case B compensates for the THz phase reversal and emits radiation with the same phase from two FMs yielding an enhanced THz pulse upon constructive interference. Overall, the THz emission exhibits a complementary behavior with static magnetization. For example, in Case A, while the two THz pulses driven by individual FMs destructively interfere and produce $\boldsymbol{THz}^{(\parallel)} = \boldsymbol{THz_1} - \boldsymbol{THz_2}$, their static magnetizations add up, given by $\boldsymbol{M}^{(\parallel)} = \boldsymbol{M_1} + \boldsymbol{M_2}$. The subscripts 1 and 2 pertain to $FM_1$ and $FM_2$, while superscripts ($\parallel$) and ($\nparallel$) denote the relative magnetization of the two ferromagnets in parallel and anti-parallel configurations, respectively. The magnetic field-dependent sweep of THz pulse amplitude measured with terahertz spintronic magnetometer (TSM)[46] is shown in Figure 2(b-e) with varying interlayer Ru thickness. As such, the comparison between the THz-H hysteresis[46,47] (TSM) and M-H hysteresis (taken from vibration spintronic magnetometer, VSM) is observed to demonstrate an exact complementary behavior across the magnetic field. We see an empirical relation where Figure 2(b-e) follows Equation 1, as given by,

$$\frac{THz_{amp}^{(\parallel)}}{THz_{amp}^{(\#)}} = \frac{M^{(\#)}}{M^{(\parallel)}} \tag{1}$$

$$\Rightarrow \frac{j_{c1}^{(\parallel)} + j_{c2}^{(\parallel)}}{j_{c1}^{(\#)} + j_{c2}^{(\#)}} = \frac{M_1 - M_2}{M_1 + M_2}$$

$$\Rightarrow \frac{\gamma_1^{total} j_{s1} - \gamma_2^{total} j_{s2}}{\gamma_1^{total} j_{s1} + \gamma_2^{total} j_{s2}} = \frac{j_{s1} - j_{s2}}{j_{s1} + j_{s2}} \tag{2}$$

In the above equations, the $THz_{amp}$ denotes the amplitude of the THz pulse, which is known to scale proportionally with the $j_c$ (refer to Supplementary Methods section for THz-H hysteresis measurement details)[46]. However, equation (2) holds true only if $\gamma_1^{total} = \gamma_2^{total}$ and highlights that the spins pumped from the FM$_1$ and FM$_2$ undergo symmetric transport processes and experience equal SCC. Subsequently, two fundamental questions emerge: 1) Is the spin-to-charge conversion (SCC) at the interface between the ferromagnet (FM) and the heavy-metal (HM) significant? 2) If so, what proportion of the overall SCC is attributed to the interface between FM/HM, and what fraction originates from the bulk HM? To address these, we focus on the transition region in white, as shown in Figure 2(b). In the white region, the external magnetic field changes the relative spin polarization of the two ferromagnets and affects the spin mobility across the interface but does not impact the SCC in bulk HM. Therefore, we analyze the spin transport around the transition region using a well-known spin-torque ferromagnetic resonance (STFMR) technique[48–50].

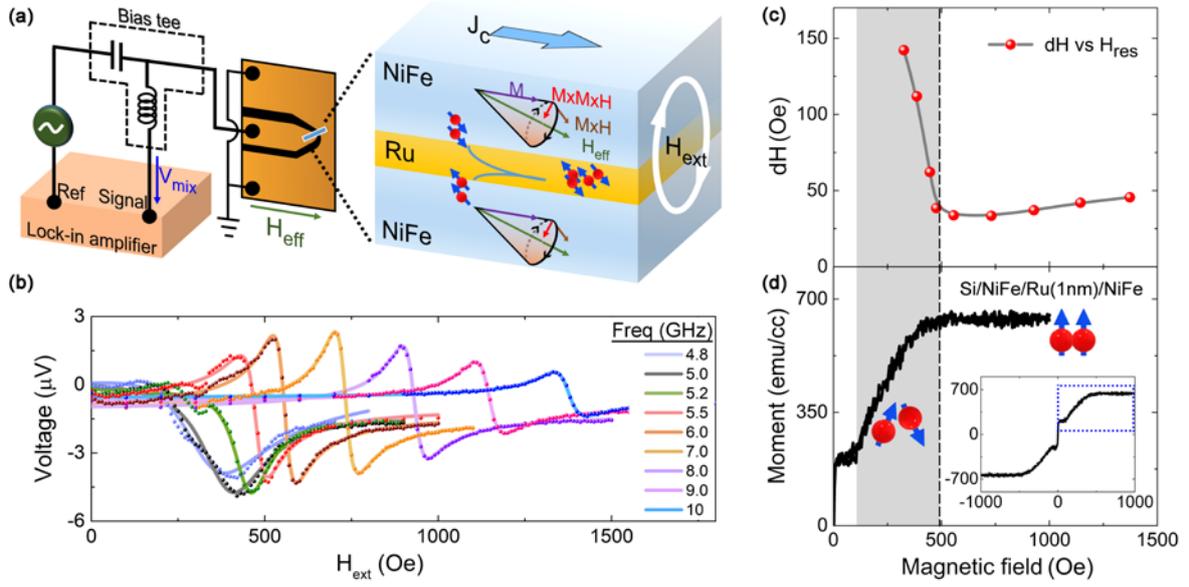

**Figure 3: Spin-Torque ferromagnetic resonance measurement (FMR) -** (a) Setup schematic as used for FMR experiment on Si/NiFe(3nm)/Ru(1nm)/NiFe(3nm) (b) FMR spectra recorded from the experiment at variable radio-frequencies between 4.8 to 10 GHz. (c) FMR spectra linewidth dependence on the resonant external magnetic field. (d) The magnetic hysteresis of the heterostructure in the positive magnetic field illustrates the transition between the two states of the pair of ferromagnets: anti-parallel and parallel states. The inset shows the magnetic hysteresis of the sample in the full range.

As shown in **Figure 3(a)**, we fabricated 5 μm broad and 25 μm long NiFe/Ru(1 nm)/NiFe strips over a silicon substrate using a photolithographic technique. The film was deposited together with the sample used for THz emission. A radio-frequency current is applied to precess the spins of the FM layer, and ferromagnetic resonance (FMR) spectra are recorded by measuring the rectified anisotropic magnetoresistance (AMR) voltage[51]. As shown in Figure 3(b), the rectified AMR voltage, given by $V_{mix}$, is measured at varying frequencies and modeled using the generalized Lorentzian curve as given by

$$V_{mix} = S * F_{sym}(H_{ext}) + A * F_{asym}(H_{ext})$$

$$F_{sym} = \frac{dH^2}{dH^2 + (H_{ext} - H_{res})^2} ; F_{asym} = \frac{dH(H_{ext} - H_{res})}{dH^2 + (H_{ext} - H_{res})^2}$$

where d$H$ is the linewidth of the resonant FMR spectra, $H_{res}$ is the resonant magnetic field, $H_{ext}$ is the applied magnetic field, S is the symmetric Lorentzian coefficient, and A is the antisymmetric Lorentzian coefficient. The FMR linewidth (d$H$) as a function of $H_{res}$ in the NiFe/Ru/NiFe structure is shown in Figure 3(c). As we progress toward decreasing $H_{res}$, it is observed that the linewidth remains relatively constant until reaching 500 Oe. However, a sharp increase in the linewidth occurs within the range of 300 Oe < $H_{res}$ < 500 Oe. Besides, a varying FMR linewidth signifies a change in the spin loss[52] within the spintronic heterostructure, and therefore the spin-to-charge conversion (SCC) within the device can be controlled by an applied magnetic field. While the SCC of the bulk heavy metal (HM) remains unaffected by the external magnetic field, it is evident that the SCC at the interface between the ferromagnet (FM) and HM is subject to change. Moreover, as we go from 500 Oe to 100 Oe, Figure 3(d) exhibits a simultaneous change in relative FM magnetization from the parallel magnetic state to the anti-parallel magnetic state. Hence, the direct evidence for the existence of interfacial spin-to-charge conversion (*i*SCC) is established by the comprehensive correlation observed as the scissor states, which are responsible for the magnetic-field-tailored spin accumulation and spin mobility as highlighted in the grey region in Figures 3(c and d).

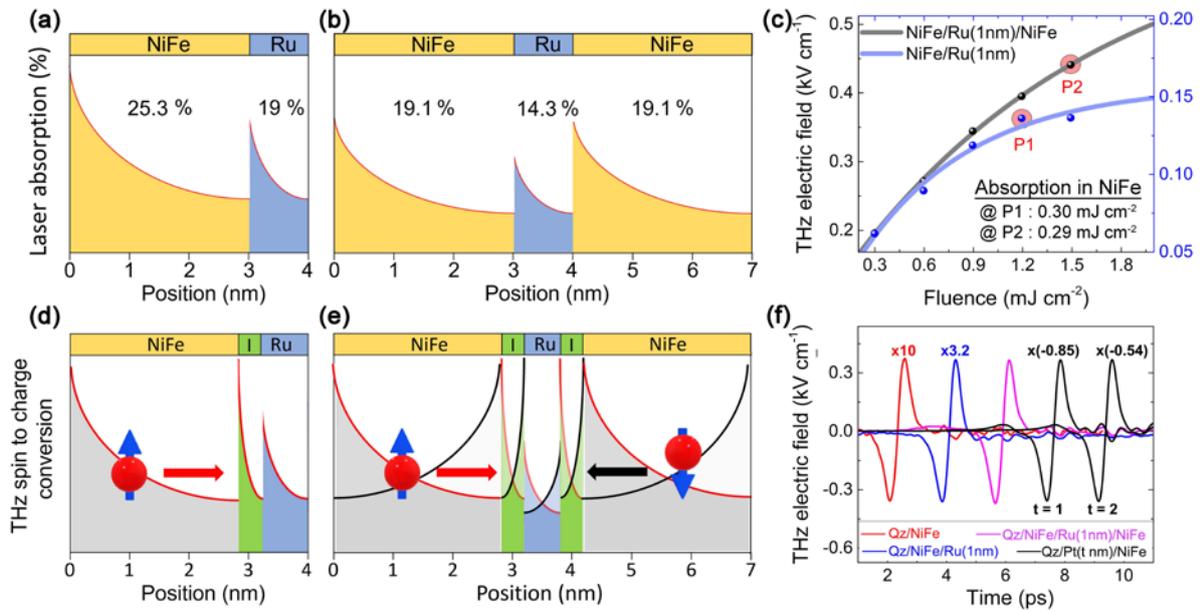

**Figure 4: Comparison between THz emission from Qz/NiFe(3nm)/Ru(1nm)/NiFe(3nm) and Qz/NiFe(3nm)/Ru(1nm) -** (a-b) Laser absorption calculation for individual NiFe and Ru layers in (a) NiFe/Ru and (b) NiFe/Ru/NiFe heterostructure (c) Fluence dependent THz pulse amplitude from the NiFe/Ru and NiFe/Ru/NiFe heterostructure. (d-e) Schematic depicting the ultrafast SCC through exponential decay of spin polarization at FM/HM interface [green area], bulk HM [blue area], and bulk FM [grey area] in (d) NiFe/Ru and (e) NiFe/Ru/NiFe heterostructure. (d-e) The red(black) exponential curve corresponds to the relaxation of forward(backward) traveling spins. (f) Emitted THz pulse from (i) Qz/NiFe (ii) Qz/NiFe/Ru(1nm) (iii) Qz/NiFe/Ru(1nm)/NiFe (iv) Qz/Pt(1nm)/NiFe (v) Qz/Pt(2nm)/NiFe.

Following the evident spin relaxation processes in NiFe/Ru interface, we also acknowledge that the STFMR possesses a key difference in the excitation energy when compared to the femtosecond pulse excitation and is not trivial to assume similar strengths of SCC[53]. Upon photoexcitation, both the ferromagnets generate ultrafast superdiffusive [41,42] spin currents, and as shown above, they experience an equal SCC within the heterostructure. Taking this into account, we proceed to deconvolute the ultrafast SCC arising in the heterostructure by comparing the terahertz emission from the two devices, NiFe/Ru(1nm)/NiFe and NiFe/Ru(1nm). Here, an equal magnetic moment was ensured using the VSM measurement, as shown in Supplementary Section S1. Further, in **Figures 4(a and b)**, we evaluate the laser absorption percentage in the spintronic heterostructures. Poynting's theorem is used to estimate the absorption allocation of the excitation laser in the individual layer based on the experimentally observed reflectance and transmittance from the samples (Refer to Supplementary Methods section regarding absorption calculation, including the application of Poynting's theorem[54]). From Figure 4(b), we find that the SAF heterostructure exhibits the same fluence absorption, 19.1%, in both the NiFe layers, suggesting the formation of equal spin current density. Besides, as shown in Figure 4(a), NiFe absorbs a higher fraction of laser fluence, 25.3%, but an equal fluence absorption is preferred for a one-to-one comparison of the ultrafast SCC within the two heterostructures. Therefore, in Figure 4(c), the THz emission from NiFe/Ru/NiFe (black spheres with an exponential curve in a solid black line) and NiFe/Ru (Blue spheres with an exponential curve in a solid blue line) is recorded with increasing incident laser fluence. We identify that if we choose to illuminate SAF heterostructure with a fluence of 1.5 mJ cm$^{-2}$, the absorption in each NiFe layer is 0.29 mJ cm$^{-2}$ (indicated as point P2), and a fluence of 1.2 mJ cm$^{-2}$ on NiFe/Ru bilayer led to 0.30 mJ cm$^{-2}$ absorption in NiFe (indicated as point P1). Hence, we investigate the THz emission from SAF heterostructure at a fluence of 1.5 mJ cm$^{-2}$ marked by P2 and the THz emission from NiFe/Ru at a fluence of 1.2 mJ cm$^{-2}$ marked by P1.

Consequently, we can assume a symmetric transport behavior in the NiFe/Ru/NiFe heterostructure, making the NiFe/Ru bilayer a suitable choice as its control sample. Figures 4(d and e) illustrate the overall SCC sites in both heterostructures. In Figure 4(d), the NiFe/Ru bilayer shows one FM/HM interface, one bulk HM site, and one bulk FM site for SCC, whereas, in Figure 4(e), NiFe/Ru/NiFe exhibits SCC from two FM/HM interfaces, one bulk HM site and two bulk FM site for each NiFe layer. Upon photoexcitation, the FM spins travel towards the Ru interlayer from both directions, as highlighted in black and red color, while the spin polarization decay exponentially. The formalism allows us to apply the symmetry arguments and compare the THz emission, as shown in Figure 4(f). Here, we observe the THz emission from SAF to be 3.2 times that from the control NiFe/Ru heterostructure (THz pulse for SAF is shown in a solid pink line at ~100 Oe when two FMs are in the anti-parallel state as shown by Case B in Figure 2(a). The THz pulse for NiFe/Ru heterostructure is shown in a solid blue line. Thus, using the general formalism of $\boldsymbol{j_c} = \gamma^{total} \boldsymbol{j_s} \times \boldsymbol{M}/|\boldsymbol{M}|$, the relation between the THz emission from the two heterostructures is given in Equation 3 as,

$$3.2 \times [j_s^{NiFe}(\gamma^{AHE} + \gamma^{Ru} + \gamma^I)] = j_s^{NiFe_1}(\gamma^{AHE} + \gamma^{I_1} + \gamma^{Ru} + \gamma^{I_2} + \gamma^{AHE})$$
$$+ j_s^{NiFe_2}(\gamma^{AHE} + \gamma^{I_1} + \gamma^{Ru} + \gamma^{I_2} + \gamma^{AHE}) \quad (3)$$

Upon following Supplementary Section S1, where an equal magnetic moment from each FM is observed, we consider $j_s^{NiFe} = j_s^{NiFe_1} = j_s^{NiFe_2}$. An equal SCC experienced by the spin yields $\gamma^I = \gamma^{I_1} = \gamma^{I_2}$. In addition, when considering the synthetic antiferromagnetic (SAF) structure, an adjustment factor is incorporated to account for the 78% THz transmission through the 3 nm NiFe layer (see Supplementary Section S5 for transmission calculation). This correction is necessary due to the positioning of the NiFe layer following the Ru layer, resulting in Equation 4 as follows:

$$3.2 \times \left[j_s^{NiFe}(\gamma^{AHE} + \gamma^{Ru} + \gamma^I)\right] = 0.78 \times \begin{bmatrix} j_s^{NiFe}(\gamma^{AHE} + \gamma^{I_1} + \gamma^{Ru} + \gamma^{I_2} + \gamma^{AHE}) \\ +j_s^{NiFe}(\gamma^{AHE} + \gamma^{I_1} + \gamma^{Ru} + \gamma^{I_2} + \gamma^{AHE}) \end{bmatrix} \quad (4)$$

Moreover, as shown in Figure 4(f), the THz emission from NiFe/Ru is found to be 3.125 times (calculated by 10/3.2) that of emission from bare NiFe, which suggests:

$$10 \times \left[j_s^{NiFe}(\gamma^{AHE})\right] = 3.2 \times j_s^{NiFe}(\gamma^{AHE} + \gamma^{I_1} + \gamma^{Ru}) \quad (5)$$

The calculation from Equations 4 and 5 returns $|\gamma^I| = 14 \times |\gamma^{Ru}|$ and indicates the SCC at the Ru interlayer to be negligible. On the contrary, a counter-argument proposes that if the *i*SCC is deemed negligible, the terahertz (THz) emission from the synthetic antiferromagnetic (SAF) structure should not exceed 2.64 times that of the control sample employed. However, our findings reveal a factor of approximately 4.10 times (calculated as 3.2 divided by 0.78), indicating a significantly higher THz emission in the SAF structure compared to the expected limit. Therefore, it can be inferred that the SAF structure primarily consists of two ferromagnet/heavy-metal (FM/HM) interface layers, with a minimal contribution from the bulk heavy metal (HM) region. To assess the strength of *i*SCC on an absolute scale, we additionally compare the terahertz (THz) emission between the SAF structure and the NiFe/Pt(*t* nm) configuration; *t*=1,2 nm, as shown by the solid black line in Figure 4(f). We know the spin diffusion length in Platinum, $\lambda_{sd}^{Pt}$ ~1.2 nm[55]. Hence, a platinum thickness of 1nm exhibits contribution that predominantly originates from the interfaces. In contrast, the NiFe/Pt(2 nm) configuration incorporates both the interface and bulk effects. From Figure 4(f), it is apparent that each NiFe/Ru interface exhibits a substantial 42.5% contribution to SCC compared to that from Platinum in NiFe/Pt(1 nm) and a 27% contribution compared to SCC from Platinum in NiFe/Pt(2 nm). These results provide compelling evidence and underscore the unexplored potential of *i*SCC, highlighting their significance in the realm of ultrafast spintronics.

In conclusion, we emphasize the FM/HM interface layer to host a rich platform for efficient spin-to-charge conversion and show a potential route to surpass the perceived saturation limit in THz emission. The experiment results highlight a giant *i*SCC at the interface, 27% as compared to that of the Platinum and pins down a fundamental prospect which has so far remained elusive. Hence, a qualitative and quantitative analysis of the interfacial *i*SCC will facilitate the development of ultrafast data processing and spintronic terahertz science. Furthermore, the results reinvigorate a broad class of study, including topological insulators, Weyl semimetals, and 2-dimensional semiconductors for their application as an interfacial SCC layer.

**Supplementary information** is linked to the online version of the paper.

**Acknowledgments**

R.S. and P.A. would like to acknowledge National Research Foundation, Singapore, for the support through NRF-CRP23-2019-0005. We also thank Prof. Lew Wen Siang and Mr. Calvin Ang Ching Ian for helping with VSM measurements.

**Author Contributions**

P.A., R.M., and R.S. conceived the project and designed the experiments. P.A. and K.D. performed all the THz measurements and experimental analysis. J.R.M and Y.F. performed the FMR measurements. P.A. and R.M. performed the FMR measurement analysis. Y.Y. provided the theoretical model. H.A. and Y.F. fabricated the spintronic emitter. All the authors analyzed and discussed the results. P.A., R.M, and R.S. wrote the manuscript with inputs from all the authors. R.S. lead the overall project.

**Data Availability Statement**

The data that support the findings of this study are available from the corresponding author upon reasonable request.

**Author Information**

Reprints and permissions information is available online. The authors declare no competing financial interests. Correspondence and requests for materials should be addressed to Dr. Ranjan Singh (ranjans@ntu.edu.sg).